\documentclass[aip,jmp,11pt,a4paper]{revtex4-1}
\usepackage{amsmath}
\usepackage{amsfonts}
\usepackage{amssymb}
\usepackage{graphicx}

\providecommand{\U}[1]{\protect \rule{.1in}{.1in}}
\newtheorem{theorem}{Theorem}

\newtheorem{corollary}{Corollary}

\newtheorem{proposition}{Proposition}

\input{tcilatex}
\begin{document}

\title{New concise upper bounds on quantum violation of general multipartite
Bell inequalities }
\author{Elena R. Loubenets}
\date{27 March 2017}

\begin{abstract}
Last years, bounds on the maximal quantum violation of \emph{general} Bell
inequalities were intensively discussed in the literature via different
mathematical tools. In the present paper, we analyze quantum violation of
general Bell inequalities via \emph{the LqHV (local quasi hidden variable)
modelling framework, }correctly reproducing\ the probabilistic description
of every quantum correlation scenario. The LqHV mathematical framework
allows us to derive for all $d$ and $N$ a new upper bound $(2d-1)^{N-1}$ on
the maximal violation by an $N$-qudit state of all general\emph{\ }Bell
inequalities, also, new upper bounds on the maximal violation by an $N$%
-qudit state of general Bell inequalities for $S$ settings per site. These
new upper bounds essentially improve all the known precise upper bounds on
quantum violation of general multipartite Bell inequalities. For some $S,$ $%
d $ and $N,$ the new upper bounds are attainable.
\end{abstract}

\maketitle

\affiliation{National Research University Higher School of Economics,\\
Moscow, 101000, Russia}

\section{Introduction}

Quantum violation of Bell inequalities\cite{1} is now used in many quantum
information tasks and is also important for the analysis of nonlocal games
strategies in computer science. The most analytically studied\cite{2, 3, 4,
5} cases of quantum violation of specific Bell inequalities refer to the
Clauser-Horne-Shimony-Holt (CHSH) inequality and the Mermin-Klyshko
inequality. It is also well known that the maximal quantum violation of 
\emph{correlation} bipartite Bell inequalities cannot\cite{6} exceed the
real Grothendiek's constant $K_{G}^{(\mathbb{R})}\in \lbrack 1.676,1.783]$
independently of a dimension of a bipartite quantum state and numbers of
settings and outcomes per site. But this is not already the case for quantum
violation of bipartite Bell inequalities on joint probabilities and last
years bounds on the maximal quantum violation of Bell inequalities were
intensively discussed in the literature\ via different mathematical tools%
\cite{7, 8, 9, 10, 11, 12, 13, 14, 15}.

To our knowledge, the maximal violation by an $N$-qudit quantum state of 
\emph{general}\cite{16} Bell inequalities for arbitrary numbers of
measurement settings and outcomes at each site admits the following upper
bounds.

\begin{itemize}
\item $N=2:$ \textrm{(a)} for an arbitrary two-qudit state -- the precise%
\cite{17} upper bound $(2d-1)$ in Eq. (64) of Ref. 9 and the precise upper
bound $2d$ in\ Proposition 5.2 of Ref. 14; \textrm{(b)} for the two-qudit
Greenberger-Horne-Zeilinger (GHZ) state -- the upper bound $Cd/\sqrt{\ln d},$
found up to a universal constant in Theorem 0.3 of Ref. 11.

\item $N\geq 3:$ \textrm{(c)} for the $N$-qudit GHZ state -- the precise
upper bound $(2^{N-1}(d-1)+1)$ in Eq. (58) of Ref. 9; \textrm{(d)} for an
arbitrary $N$-qudit state --\textrm{\ }the precise upper bound $%
(2^{N-1}d^{N-1}-2^{N-1}+1)$ in Eq. (62) of Ref. 9 and the precise upper
bound $(2d)^{N-1}$ in comments after Proposition 5.2 in Ref. 14.
\end{itemize}

In the present paper, we analyze the maximal quantum violation of general
Bell inequalities via \emph{the LqHV (local quasi hidden variable) modelling
framework, }introduced and developed in Refs. 9, 18, 19. A general
correlation scenario admits a LqHV model if and only if it is nonsignaling%
\cite{20}. Therefore, the probabilistic description of each quantum
correlation scenario admits\cite{9} the LqHV modelling. Moreover, the
probabilistic description of all projective $N$-partite joint quantum
measurements on an $N$-qudit state can be reproduced\cite{21} via the single
LqHV model specified in Ref. 13.

The LqHV mathematical framework allows us to derive a new precise upper
bound 
\begin{equation}
\left( 2d-1\right) ^{N-1}  \label{2}
\end{equation}%
on the maximal violation by an arbitrary $N$-qudit state of \emph{general\ }%
Bell inequalities for arbitrary numbers of settings and outcomes per site.
For $N=2,$ this new bound reduces to our upper bound (64) in Ref. 9. For all 
$N\geq 3,$ the new bound (\ref{2}) essentially improves all the known
precise upper bounds for general multipartite Bell inequalities, see in item
(d) above.

For the maximal quantum violation of \emph{general} Bell inequalities for $S$
settings per site, the new upper bound (\ref{2}) allows us also to improve
due to%
\begin{equation}
\left( 2\min \{d,S\}-1\right) ^{N-1}  \label{2_1}
\end{equation}%
the precise upper bound (62) in Ref. 9 for generalized $N$-partite joint
quantum measurements and due to 
\begin{eqnarray}
\min {\Large \{}d^{\frac{N-1}{2}},\text{ }3^{N-1}{\Large \}},\text{ \ \ \ \
for \ }S &=&2,  \label{2_2} \\
\min {\Large \{}d^{\frac{S(N-1)}{2}},\text{ }\left( 2\min \{d,S\}-1\right)
^{N-1}{\Large \}},\text{ \ \ \ for }S &\geq &3,  \notag
\end{eqnarray}%
the precise upper bound (19) in Ref. 13 for projective $N$-partite joint
quantum measurements. For some $d,$ $S$ and $N,$ the upper bounds (\ref{2_1}%
), (\ref{2_2}) are attainable, see in Section VI.

The main results of the present paper are formulated by theorem 1 and
corollary 1 in Section V.

\section{Preliminaries: general Bell inequalities}

In this section, we shortly recall the notion of a general Bell inequality.
The general framework for multipartite Bell inequalities for an arbitrary
number of measurement settings and any spectral type of outcomes at each
site was introduced in Ref. 22 where specific examples of Bell inequalities
are discussed in section 3.

Consider an $N$-partite correlation scenario\cite{23}, where each $n$-th of $%
N\geq 2$ parties performs $S_{n}\geq 1$ measurements with outcomes $\lambda
_{n}\in \Lambda _{n}$ of any nature and an arbitrary spectral type. We label
each measurement at $n$-th site by a positive integer $s_{n}=1,...,S_{n}$
and each $N$-partite joint measurement, induced by this correlation scenario
and with outcomes 
\begin{equation}
(\lambda _{1},\ldots ,\lambda _{N})\in \Lambda =\Lambda _{1}\times \cdots
\times \Lambda _{N}  \label{0}
\end{equation}%
by an $N$-tuple $(s_{1},...,s_{N}),$ where $n$-th component specifies a
measurement at $n$-th site. For concreteness, we denote by $\mathcal{E}%
_{S,\Lambda },$ $S=S_{1}\times \cdots \times S_{N},$ an $S_{1}\times \cdots
\times S_{N}$-setting correlation scenario with outcomes in $\Lambda $ and
by $P_{(s_{1},...,s_{N})}^{(\mathcal{E}_{S,\Lambda })}$ -- a joint
probability distribution of outcomes $(\lambda _{1},\ldots ,\lambda _{N})\in
\Lambda $ for an $N$-partite joint measurement $(s_{1},...,s_{N})$ under a
scenario $\mathcal{E}_{S,\Lambda }$.

An $N$-partite correlation scenario $\mathcal{E}_{S,\Lambda }$ is referred
to as \emph{nonsignaling} if, for any two joint measurements $%
(s_{1},...,s_{N})$ and $(s_{1}^{\prime },...,s_{N}^{\prime })$ with common
settings $s_{n_{1}},...,s_{n_{_{M}}}$ at some $1\leq n_{1}<$ $...<n_{M}\leq
N $ sites, the marginal probability distributions of distributions $%
P_{(s_{1},...,s_{N})}^{(\mathcal{E}_{S,\Lambda })}$ and $P_{(s_{1}^{\prime
},...,s_{N}^{\prime })}^{(\mathcal{E}_{S,\Lambda })}$, describing
measurements at sites $1\leq n_{1}<$ $...<n_{M}\leq N$, coincide. For
details, see section 3 in Ref. 24.

For a correlation scenario $\mathcal{E}_{S,\Lambda },$ consider a linear
combination 
\begin{eqnarray}
\mathcal{B}_{\Phi _{S,\Lambda }}^{(\mathcal{E}_{S,\Lambda })}
&=&\sum_{s_{1},...,s_{_{N}}}\left\langle f_{(s_{1},...,s_{N})}(\lambda
_{1},\ldots ,\lambda _{N})\right\rangle _{\mathcal{E}_{S,\Lambda }},
\label{1.1} \\
\Phi _{S,\Lambda } &=&\{f_{(s_{1},...,s_{N})}:\Lambda \rightarrow \mathbb{R}%
\mid s_{n}=1,...,S_{n},\text{ \ }n=1,...,N\},  \notag
\end{eqnarray}%
of averages (expectations) of the most general form%
\begin{eqnarray}
&&\left\langle f_{(s_{1},...,s_{N})}(\lambda _{1},\ldots ,\lambda
_{N})\right\rangle _{\mathcal{E}_{S,\Lambda }}  \label{1.2} \\
&=&\int\limits_{\Lambda }f_{(s_{1},...,s_{N})}(\lambda _{1},\ldots ,\lambda
_{N})P_{(s_{1},...,s_{N})}^{(\mathcal{E}_{S,\Lambda })}\left( \mathrm{d}%
\lambda _{1}\times \cdots \times \mathrm{d}\lambda _{N}\right) ,  \notag
\end{eqnarray}%
specified for each joint measurement $(s_{1},...,s_{N})$ by a bounded
real-valued function $f_{(s_{1},...,s_{N})}(\cdot )$ of outcomes $\left(
\lambda _{1},\ldots ,\lambda _{N}\right) \in \Lambda $ at all $N$ sites.

Depending on a choice of a function $f_{(s_{1},...,s_{N})}$ for a joint
measurement $(s_{1},...,s_{N})$, an average (\ref{1.2}) may refer either to
the joint probability of events observed at $M\leq N$ sites or, in case of
real-valued outcomes, for example, to the expectation 
\begin{equation}
{\Large \langle }\lambda _{1}^{(s_{1})}\cdot \ldots \cdot \lambda
_{n_{M}}^{(s_{n_{M}})}{\Large \rangle }_{\mathcal{E}_{S,\Lambda
}}=\int\limits_{\Lambda }\lambda _{1}\cdot \ldots \cdot \lambda
_{n_{M}}P_{(s_{1},...,s_{N})}^{(\mathcal{E}_{S,\Lambda })}\left( \mathrm{d}%
\lambda _{1}\times \cdots \times \mathrm{d}\lambda _{N}\right)  \label{1.3}
\end{equation}%
of the product of outcomes observed at $M\leq N$ sites or may have a more
complicated form. In quantum information, the product expectation (\ref{1.3}%
) is referred to as a correlation function. For $M=N,$ a correlation
function is called full.

The probabilistic description of an arbitrary correlation scenario $\mathcal{%
E}_{S,\Lambda }$ admits\cite{25} \emph{a LHV (local hidden variable) model}
if all its joint probability distributions%
\begin{equation}
\left \{ P_{(s_{1},...,s_{N})}^{(\mathcal{E}_{S,\Lambda })},\text{ }%
s_{1}=1,...,S_{n},...,s_{N}=1,...,S_{N}\right \}  \label{1.4}
\end{equation}%
admit the representation%
\begin{eqnarray}
&&P_{(s_{1},...,s_{N})}^{(\mathcal{E}_{S,\Lambda })}\left( \mathrm{d}\lambda
_{1}\times \cdots \times \mathrm{d}\lambda _{N}\right)  \label{1.5} \\
&=&\dint \limits_{\Omega }P_{1,s_{1}}(\mathrm{d}\lambda _{1}|\omega )\cdot
\ldots \cdot P_{N,s_{_{N}}}(\mathrm{d}\lambda _{N}|\omega )\nu _{\mathcal{E}%
_{S,\Lambda }}(\mathrm{d}\omega )  \notag
\end{eqnarray}%
via a single probability distribution $\nu _{\mathcal{E}_{S,\Lambda }}(%
\mathrm{d}\omega )$ of some variables $\omega \in \Omega $ and conditional
probability distributions $P_{n,s_{n}}(\mathrm{\cdot }|\omega ),$ referred
to as "local" in the sense that each $P_{n,s_{n}}(\mathrm{\cdot }|\omega )$
at $n$-th site depends only on the corresponding measurement $%
s_{n}=1,...,S_{n}$ at this site.

Let a correlation scenario $\mathcal{E}_{S,\Lambda }$ admit an LHV model.
Then a linear combination (\ref{1.1}) of its averages (\ref{1.2}) satisfies
the tight\cite{26} LHV constraint (see Theorem 1 in Ref. 22): 
\begin{equation}
\mathcal{B}_{\Phi _{S,\Lambda }}^{\inf }\leq \mathcal{B}_{\Phi _{S,\Lambda
}}^{(\mathcal{E}_{S,\Lambda })}{\Large |}_{_{lhv}}\leq \mathcal{B}_{\Phi
_{S,\Lambda }}^{\sup }  \label{1.6}
\end{equation}%
with the LHV constants%
\begin{eqnarray}
\mathcal{B}_{\Phi _{S,\Lambda }}^{\sup } &=&\sup_{\lambda _{n}^{(s_{n})}\in
\Lambda _{n},\forall s_{n},\forall n}\text{ }%
\sum_{s_{1},...,s_{_{N}}}f_{(s_{1},...,s_{N})}(\lambda _{1}^{(s_{1})},\ldots
,\lambda _{N}^{(s_{N})}),  \label{1.7} \\
\mathcal{B}_{\Phi _{S,\Lambda }}^{\inf } &=&\inf_{\lambda _{n}^{(s_{n})}\in
\Lambda _{n},\forall s_{n},\forall n}\text{ }%
\sum_{s_{1},...,s_{_{N}}}f_{(s_{1},...,s_{N})}(\lambda _{1}^{(s_{1})},\ldots
,\lambda _{N}^{(s_{N})}).  \notag
\end{eqnarray}%
From (\ref{1.6}), it follows that, in the LHV\ case, 
\begin{equation}
\left\vert \text{ }\mathcal{B}_{\Phi _{S,\Lambda }}^{(\mathcal{E}_{S,\Lambda
})}{\Large |}_{_{lhv}}\right\vert \leq \mathcal{B}_{\Phi _{S,\Lambda
}}^{lhv}=\max \left\{ \left\vert \mathcal{B}_{\Phi _{S,\Lambda }}^{\sup
}\right\vert ,\left\vert \mathcal{B}_{\Phi _{S,\Lambda }}^{\inf }\right\vert
\right\} .  \label{1.8}
\end{equation}

Note that some of the LHV inequalities in (\ref{1.6}) may be fulfilled for a
wider (than LHV) class of correlation scenarios. This is, for example, the
case for the LHV\ constraints on joint probabilities following explicitly
from nonsignaling of probability distributions. Moreover, some of the LHV
inequalities in (\ref{1.6}) may be simply trivial, i. e. fulfilled for all
correlation scenarios, not necessarily nonsignaling.

\emph{Each of the tight LHV inequalities in (\ref{1.6}) that may be violated
under a non-LHV scenario is referred to as a Bell (or Bell-type) inequality. 
}

\section{Quantum violation}

Let, under a correlation scenario with $S_{n}$ measurement settings and
outcomes $\lambda _{n}\in \Lambda _{n}$ at each $n$-th site, every $N$%
-partite joint measurement $(s_{1},...,s_{N})$ be performed on a quantum
state $\rho $ on a complex Hilbert space $\mathcal{H}_{1}\otimes \cdots
\otimes \mathcal{H}_{N}$ and be described by the joint probability
distribution 
\begin{equation}
\mathrm{tr}[\rho \{ \mathrm{M}_{1,s_{1}}(\mathrm{d}\lambda _{1})\otimes
\cdots \otimes \mathrm{M}_{N,s_{n}}(\mathrm{d}\lambda _{N})\}],  \label{1.11}
\end{equation}%
where each $\mathrm{M}_{n,s_{n}}(\mathrm{d}\lambda _{n})$ is a normalized
positive operator-valued (\emph{POV}) measure, representing on a complex
Hilbert space $\mathcal{H}_{n}$ a quantum measurement $s_{n}$ at $n$-th
site. For a POV measure $\mathrm{M}_{n,s_{n}}$, all its values $\mathrm{M}%
_{n,s_{n}}(F_{n}),$ $F_{n}\subseteq \Lambda _{n},$ are positive operators on 
$\mathcal{H}_{n}$ and $\mathrm{M}_{n,s_{n}}(\Lambda _{n})=\mathbb{I}_{%
\mathcal{H}_{n}}.$ For concreteness, we specify this $S_{1}\times \cdots
\times S_{N}$-setting quantum correlation scenario by symbol $\mathcal{E}_{%
\mathrm{M}_{S,\Lambda }}^{(\rho )}$ where 
\begin{eqnarray}
\mathrm{M}_{S,\Lambda } &:&=\left \{ \mathrm{M}_{n,s_{n}},\text{ }s_{n}=1,..%
\mathbf{.},S_{n},\text{ }n=1,...,N\right \} ,  \label{1.12} \\
S &=&S_{1}\times \cdots \times S_{N},\text{ \ \ }\Lambda =\Lambda _{1}\times
\cdots \times \Lambda _{N},  \notag
\end{eqnarray}%
is a collection of POV measures (\ref{1.11}) at all $N$-sites.

As it is well known, the probabilistic description of a quantum correlation
scenario $\mathcal{E}_{\mathrm{M}_{S,\Lambda }}^{(\rho )}$ does not need to
admit a LHV model. Therefore, under correlation scenarios on an $N$-partite
quantum state $\rho ,$ Bell inequalities (\ref{1.6}) may be violated and the
parameter\cite{9} 
\begin{equation}
\mathrm{\Upsilon }_{S_{1}\times \cdots \times S_{N}}^{(\rho ,\Lambda
)}=\sup_{_{\Phi _{S,\Lambda },\text{ }\mathrm{M}_{S,\Lambda }}}\frac{1}{%
\mathcal{B}_{\Phi _{S,\Lambda }}^{lhv}}\left \vert \mathcal{B}_{\Phi
_{S,\Lambda }}^{(\mathcal{E}_{\mathrm{M}_{S,\Lambda }}^{(\rho )})}\right
\vert \geq 1  \label{1.15-1}
\end{equation}%
specifies the maximal violation by an $N$-partite state $\rho $ of general
Bell inequalities for $S_{1}\times \cdots \times S_{N}$-setting correlation
scenarios with outcomes $(\lambda _{1},...,\lambda _{N})$ $\in \Lambda
_{1}\times \cdots \times \Lambda _{N}=\Lambda $ at $N$ sites.

For an $N$-partite quantum state $\rho ,$ the parameter 
\begin{equation}
\mathrm{\Upsilon }_{S_{1}\times \cdots \times S_{N}}^{(\rho )}=\sup_{\Lambda
}\mathrm{\Upsilon }_{S_{1}\times \cdots \times S_{N}}^{(\rho ,\Lambda )}\geq
1  \label{1.15}
\end{equation}%
gives the maximal violation of general $S_{1}\times \cdots \times S_{N}$%
-setting Bell inequalities for an arbitrary outcome set $\Lambda _{n}$ at
each $n$-th site while the parameter 
\begin{equation}
\mathrm{\Upsilon }_{\rho }=\sup_{S_{1},...,S_{N}}\mathrm{\Upsilon }%
_{S_{1}\times \cdots \times S_{N}}^{(\rho )}\geq 1  \label{1.16}
\end{equation}%
-- the maximal violation of \emph{all} general Bell inequalities.

\section{Analytical upper bound}

We recall that, for every state $\rho $ on a complex Hilbert space $\mathcal{%
H}_{1}\otimes \cdots \otimes \mathcal{H}_{N}$ and arbitrary positive
integers $S_{1},...,S_{N}\geq 1,$ there exists\cite{27} an $S_{1}\times
\cdots \times S_{N}$-setting source operator $T_{S_{1}\times \cdots \times
S_{N}}^{(\rho )}$ -- a self-adjoint trace class operator on the space 
\begin{equation}
(\mathcal{H}_{1})^{\otimes S_{1}}\otimes \cdots \otimes (\mathcal{H}%
_{N})^{\otimes S_{N}},  \label{3}
\end{equation}%
satisfying the relation%
\begin{align}
& \mathrm{tr}\left[ T_{_{S_{1}\times \cdots \times S_{N}}}^{(\rho )}\left \{ 
\mathbb{I}_{\mathcal{H}_{1}^{\otimes k_{1}}}\otimes X_{1}\otimes \mathbb{I}_{%
\mathcal{H}_{1}^{\otimes (S_{1}-1-k_{1})}}\otimes \cdots \otimes \mathbb{I}_{%
\mathcal{H}_{N}^{\otimes k_{N}}}\otimes X_{N}\otimes \mathbb{I}_{\mathcal{H}%
_{1}^{\otimes (S_{N}-1-k_{N})}}\right \} \right]  \label{3-1} \\
& =\mathrm{tr}\left[ \rho \left \{ X_{1}\otimes \cdots \otimes X_{N}\right
\} \right] ,  \notag \\
k_{1}& =0,...,(S_{1}-1),...,k_{N}=0,...,(S_{N}-1),  \notag
\end{align}%
for all bounded linear operators $X_{1},...,X_{N}$ on Hilbert spaces $%
\mathcal{H}_{1},....,\mathcal{H}_{N},$ respectively. Here, we set $\mathbb{I}%
_{\mathcal{H}_{n}^{\otimes k}}\otimes X_{n}\mid _{_{k=0}}$ $=X_{n}\otimes 
\mathbb{I}_{\mathcal{H}_{n}^{\otimes k}}\mid _{_{k=0}}$ $=X_{n}.$

Due to its definition (\ref{3-1}), an $S_{1}\times \cdots \times S_{N}$%
-setting source operator $T_{S_{1}\times \cdots \times S_{N}}^{(\rho )}$
constitutes a self-adjoint trace class dilation of a state $\rho $ to the
complex Hilbert space (\ref{3}). Clearly, $T_{_{1\times \cdots \times
1}}^{(\rho )}\equiv \rho $ and \textrm{tr}$[T_{_{S_{1}\times \cdots \times
S_{N}}}^{(\rho )}]=1.$

The analytical bound (53) in Theorem 3 of Ref. 9, derived via the LqHV
modeling framework\cite{9, 18}, implies the following statement.

\begin{proposition}
Under all generalized $N$-partite joint quantum measurements, the maximal
violation (\ref{1.15}) by a state $\rho $ of general $S_{1}\times \cdots
\times S_{N}$-setting Bell inequalities satisfies the relation%
\begin{equation}
1\leq \mathrm{\Upsilon }_{S_{1}\times \cdots \times S_{N}}^{(\rho )}\leq
\inf_{T_{S_{1}\times \cdots \times \underset{\overset{\uparrow }{n}}{1}%
\times \cdots \times S_{N}}^{(\rho )},\text{ }\forall n}\text{ }%
||T_{S_{1}\times \cdots \times \underset{\overset{\uparrow }{n}}{1}\times
\cdots \times S_{N}}^{(\rho )}\text{ }||_{cov}  \label{4}
\end{equation}%
where: (i) infimum is taken over all source operators $T_{S_{1}\times \cdots
\times \underset{\overset{\uparrow }{n}}{1}\times \cdots \times
S_{N}}^{(\rho )}$ with only one setting at some $n$-th site and over all
sites $n=1,...,N;$ (ii) notation $\left \Vert \cdot \right \Vert _{cov}$
means the covering norm -- a new norm introduced for self-adjoint trace
class operators by relation (11) in Ref. 9.
\end{proposition}

By Lemma 1 in Ref. 9, for every self-adjoint trace class operator $W$ on a
tensor product Hilbert space $\mathcal{G}_{1}\otimes \mathcal{\cdots }%
\otimes \mathcal{G}_{m},$ \emph{its covering norm} $\left \Vert
W\right
\Vert _{cov}$ satisfies the relation%
\begin{equation}
\left \vert \mathrm{tr}\left[ W\right] \right \vert \leq \left \Vert W\right
\Vert _{cov}\leq \left \Vert W\right \Vert _{1},  \label{5}
\end{equation}%
where $\left \Vert \cdot \right \Vert _{1}$ is the trace norm and the
equality $\left \Vert W\right \Vert _{cov}=\left \vert \mathrm{tr}\left[ W%
\right] \right \vert $ is true if a self-adjoint trace class operator $W$ is 
\emph{tensor positive}, that is, satisfies the relation\cite{29}%
\begin{equation}
\mathrm{tr}\left[ W\{X_{1}\otimes \cdots \otimes X_{m}\} \right] \geq 0
\label{6}
\end{equation}%
for all positive bounded operators $X_{j}$ on $\mathcal{G}_{j},$ $j=1,...,m$%
. Every positive trace class operator is tensor positive but not vice versa.
For example, the permutation (flip) operator $V_{d}(\psi _{1}\otimes \psi
_{2}):=\psi _{2}\otimes \psi _{1},$ $\psi _{1},\psi _{2}\in \mathbb{C}^{d},$
on $\mathbb{C}^{d}\otimes \mathbb{C}^{d}$ is tensor positive but is not
positive. Its trace norm is $\left \Vert V_{d}\right \Vert _{1}=d^{2}$ while
the covering norm $\left \Vert V_{d}\right \Vert _{cov}=d.$

For every source operator $T_{S_{1}\times \cdots \times S_{N}}^{(\rho )}$,
its trace $\mathrm{tr}$[$T_{S_{1}\times \cdots \times S_{N}}^{(\rho )}]=1,$
so that, by (\ref{5}), $||T_{S_{1}\times \cdots \times S_{N}}^{(\rho
)}||_{cov}\geq 1$ and is equal to one: $||T_{S_{1}\times \cdots \times
S_{N}}^{(\rho )}$ $||_{cov}=1$ if a source operator $T_{S_{1}\times \cdots
\times S_{N}}^{(\rho )}$ is tensor positive.

This and relation (\ref{4}) imply that if, for an $N$-partite state $\rho ,$
tensor positive source operators $T_{S_{1}\times \cdots \times \underset{%
\overset{\uparrow }{n}}{1}\times \cdots \times S_{N}}^{(\rho )}$, $%
n=1,...,N, $ exist for all integers $S_{1},\ldots ,S_{N}\geq 1,$ then the
maximal violation (\ref{1.16}) by an $N$-partite state $\rho $ of general
Bell inequalities for arbitrary numbers of settings and any spectral type of
outcomes at each site is equal to one: $\mathrm{\Upsilon }_{\rho }=1$, so
that this $N$-partite quantum state $\rho $ is local in the sense that it
satisfies all general Bell inequalities.

Examples of nonseparable $N$-partite quantum states that have tensor
positive source operators $T_{S_{1}\times \cdots \times \underset{\overset{%
\uparrow }{n}}{1}\times \cdots \times S_{N}}^{(\rho )},$\ $n=1,...,N,$\ for
all integers $S_{1},\ldots ,S_{N}\geq 1$\ (and are, therefore, local) are
presented in Ref. 30.

\section{New numerical upper bounds}

Let $\rho _{d,N}$ be an arbitrary $N$-qudit quantum state on $\mathcal{H}%
^{\otimes N},$ where $\dim \mathcal{H=}d<\infty $. In order to evaluate via
the analytical bound (\ref{4}) the maximal violation $\mathrm{\Upsilon }%
_{S_{1}\times \cdots \times S_{N}}^{(\rho _{d,N})}$ by an $N$-qudit state $%
\rho _{d,N}$ of all general $S_{1}\times \cdots \times S_{N}$-setting Bell
inequalities (\ref{1.6}), we need to present at least one source operator $%
T_{S_{1}\times \cdots \times \underset{\overset{\uparrow }{n}}{1}\times
\cdots \times S_{N}}^{(\rho _{d,N})}.$ We first consider the case of a pure
state and then, by convexity, extend our result to an arbitrary $\rho _{d,N}$%
.

A pure $N$-qudit state $|\psi _{d,N}\rangle \langle \psi _{d,N}|$ admits the
decomposition 
\begin{equation}
|\psi _{d,N}\rangle \langle \psi _{d,N}|\text{ }=\sum \varsigma
_{mj...k}\varsigma _{m_{1}j_{1}...k_{1}}^{\ast }|e_{m}^{(1)}\rangle \langle
e_{m_{1}}^{(1)}|\otimes |e_{j}^{(2)}\rangle \langle e_{j_{1}}^{(2)}|\otimes
\cdots \otimes |e_{k}^{(N)}\rangle \langle e_{k_{1}}^{(N)}|,  \label{7}
\end{equation}%
where $\sum_{m,j,...,k}\left \vert \varsigma _{mj...k}\right \vert ^{2}=1$
and ${\large \{}e_{m}^{(n)}\in \mathcal{H},m=1,...,d{\large \}},$ $%
n=1,...,N, $ are orthonormal bases in $\mathcal{H}$. Introducing the
normalized vectors 
\begin{eqnarray}
\phi _{j...k} &=&\frac{1}{\beta _{j...k}}\sum_{m}\varsigma
_{mj...k}e_{m}^{(1)}\in \mathcal{H},\ \ \ \ \left \Vert \phi _{j...k}\right
\Vert =1,  \label{8} \\
\beta _{j...k} &=&\left( \dsum \limits_{m}|\varsigma _{mj...k}|^{2}\right)
^{1/2},\text{ \ \ }\sum_{\underset{N-1}{\underbrace{j,...,k}}}\beta
_{j...k}^{2}=1,  \notag
\end{eqnarray}%
we rewrite (\ref{7}) in the form%
\begin{equation}
|\psi _{d,N}\rangle \langle \psi _{d,N}|\text{ }=\sum \beta _{j...k}\beta
_{j_{1}...k_{1}}|\phi _{j...k}\rangle \langle \phi _{j_{1}...k_{1}}|\otimes
|e_{j}^{(2)}\rangle \langle e_{j_{1}}^{(2)}|\otimes \cdots \otimes
|e_{k}^{(N)}\rangle \langle e_{k_{1}}^{(N)}|.  \label{9}
\end{equation}

In view of decomposition (\ref{9}), let us introduce on the Hilbert space 
\begin{equation}
\mathcal{H}\otimes \mathcal{H}^{\otimes S_{2}}\otimes \cdots \otimes 
\mathcal{H}^{\otimes S_{N}}  \label{10}
\end{equation}%
the self-adjoint operator%
\begin{equation}
T_{1\times S_{2}\times \cdots \otimes S_{N}}^{(\psi _{d,N})}=\sum \beta
_{j...k}\beta _{j_{1}...k_{1}}|\phi _{j...k}\rangle \langle \phi
_{j_{1}...k_{1}}|\otimes W_{jj_{1}}^{(2,S_{2})}\otimes \cdots \otimes
W_{kk_{1}}^{(N,S_{N})},  \label{11}
\end{equation}%
where $W_{jj}^{(n,S_{n})}=\left( |e_{j}^{(n)}\rangle \langle
e_{j}^{(n)}|\right) ^{\otimes S_{n}}$ and 
\begin{align}
2W_{jj_{1}}^{(n,S_{n})}\mathbf{|}_{j\neq j_{1}}& =\frac{\left(
|e_{j}^{(n)}+e_{j_{1}}^{(n)}\rangle \langle
e_{j}^{(n)}+e_{j_{1}}^{(n)}|\right) ^{\otimes S_{n}}}{2^{S_{n}}}-\frac{%
\left( |e_{j}^{(n)}-e_{j_{1}}^{(n)}\rangle \langle
e_{j}^{(n)}-e_{j_{1}}^{(n)}|\right) ^{\otimes S_{n}}}{2^{S_{n}}}  \label{12}
\\
& +i\frac{\left( |e_{j}^{(n)}+ie_{j_{1}}^{(n)}\rangle \langle
e_{j}^{(n)}+ie_{j_{1}}^{(n)}|\right) ^{\otimes S_{n}}}{2^{S_{n}}}-i\frac{%
\left( |e_{j}^{(n)}-ie_{j_{1}}^{(n)}\rangle \langle
e_{j}^{(n)}-ie_{j_{1}}^{(n)}|\right) ^{\otimes S_{n}}}{2^{S_{n}}}  \notag
\end{align}%
are operators on $\mathcal{H}^{\otimes S_{n}}$ invariant with respect to
permutations of spaces $\mathcal{H}$ in $\mathcal{H}^{\otimes S_{n}}$ and
satisfying the relations%
\begin{equation}
\left( W_{jj_{1}}^{(n,S_{n})}\right) ^{\ast }=W_{j_{1}j}^{(n,S_{n})},\text{
\ \ \ \ \ }\mathrm{tr}_{\mathcal{H}^{\otimes (S_{n}-1)}}\left[
W_{jj_{1}}^{(n,S_{n})}\right] =|e_{j}^{(n)}\rangle \langle e_{j_{1}}^{(n)}|.
\end{equation}%
It is easy to verify that the partial trace 
\begin{equation}
\mathrm{tr}_{\mathcal{H}^{\otimes (S_{2}-1)}\otimes \cdots \otimes \mathcal{H%
}^{\otimes (S_{N}-1)}}\left[ T_{1\times S_{2}\times \cdots \times
S_{N}}^{(\psi _{d,N})}\right] =|\psi _{d,N}\rangle \langle \psi _{d,N}|.
\label{14}
\end{equation}%
Therefore, the self-adjoint operator $T_{1\times S_{2}\times \cdots \times
S_{N}}^{(\psi _{d,N})}$ constitutes a $1\times S_{2}\times \cdots \times
S_{N}$-setting source operator for a pure state $|\psi _{d,N}\rangle \langle
\psi _{d,N}|$.

Evaluating due to relation (\ref{5}) the covering norm of the source
operator (\ref{11})%
\begin{equation}
\left \Vert T_{1\times S_{2}\times \cdots \times S_{N}}^{(\psi
_{d,N})}\right \Vert _{cov}\leq 1+\sum_{\substack{ (j,...,k)  \\ \neq
(j_{1},...,k_{1})}}\beta _{j...k}\beta _{j_{1}...k_{1}}\text{ }\left \{
\delta _{jj_{1}}+2(1-\delta _{jj_{1}})\right \} \times \cdots \times \left
\{ \delta _{kk_{1}}+2(1-\delta _{kk_{1}})\right \}  \label{16}
\end{equation}%
and taking into account that $\sum_{\underset{N-1}{\underbrace{j,...,k}}%
}\beta _{j...k}^{2}=1$ and 
\begin{equation}
\sum_{\substack{ j,...,l  \\ \underset{m}{\underbrace{r\neq r_{1},...,k\neq
k_{1}}}}}\beta _{j...lr...k}\beta _{j..lr_{1}...k_{1}}\leq \sum_{\substack{ %
j,...,l  \\ \underset{m}{\underbrace{r\neq r_{1},...k\neq k_{1}}}}}\frac{1}{2%
}\left( \beta _{j...lr...k}^{2}+\beta _{j...lr_{1}...k_{1}}^{2}\right) \leq
(d-1)^{m},  \label{17}
\end{equation}%
also, similar relations for $m$ non-equal pairs of indices standing at
arbitrary places in the sum in (\ref{16}), we derive%
\begin{equation}
\left \Vert T_{1\times S_{2}\times \cdots \times S_{N}}^{(\psi
_{d,N})}\right \Vert _{cov}\leq \sum_{m=0}^{N-1}\binom{N-1}{m}%
2^{m}(d-1)^{m}=(2d-1)^{N-1},  \label{18}
\end{equation}%
where $\binom{N-1}{m}$ are the binomial coefficients.

From (\ref{4}), (\ref{18}) it follows that 
\begin{equation}
\mathrm{\Upsilon }_{S_{1}\times \cdots \times S_{N}}^{(\psi _{d,N})}\leq
(2d-1)^{N-1}  \label{19}
\end{equation}%
for arbitrary numbers $S_{1},...,S_{N}$ of settings at all $N$ sites. By
convexity, this upper bound is extended to an arbitrary state $\rho _{d,N}$.

Taking into account (\ref{1.16}) and incorporating also the upper bound Eq.
(58) of Ref. 9 on the maximal violation of general Bell inequalities by the $%
N$-qudit GHZ state 
\begin{equation}
\frac{1}{\sqrt{d}}\sum_{m=1}^{d}|e_{m}\rangle ^{\otimes N},  \label{19.1}
\end{equation}%
we have.

\begin{theorem}
For an arbitrary $N$-qudit state $\rho _{d,N}$, the maximal violation $%
\mathrm{\Upsilon }_{\rho _{d,N}}$ of general Bell inequalities for arbitrary
numbers of settings and outcomes at each site admits the upper bound%
\begin{equation}
\mathrm{\Upsilon }_{\rho _{d,N}}\leq (2d-1)^{N-1}  \label{20}
\end{equation}%
under all generalized $N$-partite joint quantum measurements. For the $N$%
-qudit GHZ state (\ref{19.1}), the maximal violation of general Bell
inequalities is upper bounded by%
\begin{equation}
\mathrm{\Upsilon }_{\rho _{ghz,d,N}}\leq 2^{N-1}(d-1)+1.
\end{equation}%
\smallskip
\end{theorem}

Due to the new upper bound (\ref{20}), the upper bound (62) in Ref. 9 for
generalized quantum measurements and the upper bound (19) in Ref.13 for
projective parties' measurements, we have the following corollary of Theorem
1.

\begin{corollary}
For an arbitrary $N$-qudit state $\rho _{d,N},$ the maximal violation $%
\mathrm{\Upsilon }_{S\times \cdots \times S}^{(\rho _{d,N})}$ of general
Bell inequalities for $S$ settings and an arbitrary number of outcomes at
each site satisfies the relation 
\begin{equation}
\mathrm{\Upsilon }_{S\times \cdots \times S}^{(\rho _{d,N})}\leq \left(
2\min \{d,S\}-1\right) ^{N-1}  \label{23}
\end{equation}%
under all generalized N-partite joint quantum measurements and the relation%
\begin{eqnarray}
\Upsilon _{2\times \cdots \times 2}^{(\rho _{d,N})} &\leq &\min {\Large \{}%
d^{\frac{N-1}{2}},\text{ }3^{N-1}{\Large \}},\text{ \ \ \ \ for \ }S=2,
\label{24} \\
\Upsilon _{S\times \cdots \times S}^{(\rho _{d,N})} &\leq &\min {\Large \{}%
d^{\frac{S(N-1)}{2}},\text{ }\left( 2\min \{d,S\}-1\right) ^{N-1}{\Large \}},%
\text{ \ \ \ for }S\geq 3,  \notag
\end{eqnarray}%
under projective N-partite joint quantum measurements.
\end{corollary}

\section{Discussion}

For $N=d=S=2,$ the upper bound in (\ref{24}) gives $\sqrt{2}$ and, in view
of the Cirel'son bound\cite{2}, is attained on the CHSH inequality.

For $d=S=2,$ $N\geq 3,$ the upper bound in (\ref{24}) is equal to $2^{\frac{%
N-1}{2}}$ and, due to the results in Refs. 4, 5, it is attained on the
Mermin--Klyshko inequality. The latter implies that, for \emph{projective} $%
N $-partite joint quantum measurements, the Mermin-Klyshko inequality gives
the maximal quantum violation not only among all Bell inequalities on full
correlation functions (as it was proved in Ref. 4) but also among all \emph{%
general }$N$-partite\emph{\ }Bell inequalities for two settings and two
outcomes per site.

\emph{Concerning the attainability of the term }$3^{N-1}$ \emph{in the upper
bound }$\min {\Large \{}d^{\frac{N-1}{2}},$ $3^{N-1}{\Large \}}$ \emph{in} 
\emph{(\ref{24}). }From Eq. (48) in Ref. 9 and relation $\mathrm{\Upsilon }%
_{S_{1}\times \cdots \times S_{N}}^{(\rho ,\text{ }\Lambda )}\leq \mathrm{%
\Upsilon }_{S_{1}\times \cdots \times S_{N}}^{(\rho )}$ for violation
parameters in (\ref{1.15-1}), (\ref{1.15}), it follows that, for quantum
correlation scenarios $\mathcal{E}_{\mathrm{M}_{S,\Lambda }}^{(\rho )}$ on
an $N$-partite state $\rho $, the quantum analogs of $S_{1}\times \cdots
\times S_{N}$-setting Bell inequalities (\ref{1.6}) admit the bounds:%
\begin{align}
& \mathcal{B}_{\Phi _{S,\Lambda }}^{\inf }-\frac{\mathrm{\Upsilon }%
_{S_{1}\times \cdots \times S_{N}}^{(\rho )}-1}{2}(\mathcal{B}_{\Phi
_{S,\Lambda }}^{\sup }-\mathcal{B}_{\Phi _{S,\Lambda }}^{\inf })  \label{26}
\\
& \leq \mathcal{B}_{\Phi _{S,\Lambda }}^{(\mathcal{E}_{\mathrm{M}_{S,\Lambda
}}^{(\rho )})}  \notag \\
& \leq \mathcal{B}_{\Phi _{S,\Lambda }}^{\sup }+\frac{\mathrm{\Upsilon }%
_{S_{1}\times \cdots \times S_{N}}^{(\rho )}-1}{2}(\mathcal{B}_{\Phi
_{S,\Lambda }}^{\sup }-\mathcal{B}_{\Phi _{S,\Lambda }}^{\inf }),  \notag
\end{align}%
where $\mathrm{\Upsilon }_{S_{1}\times \cdots \times S_{N}}^{(\rho )}$ is
the maximal violation (\ref{1.15}) by an $N$-partite state $\rho $ of
general $S_{1}\times \cdots \times S_{N}$-setting Bell inequalities.

Consider the Zohren-Gill (ZG) inequalities\cite{31} on joint probabilities%
\begin{equation}
1\leq \mathcal{B}_{zg}|_{_{_{_{lhv}}}}\leq 2,  \label{27}
\end{equation}%
constituting the Bell inequalities for the bipartite case with two settings $%
(N=S=2)$ and $d$ outcomes at each site.

In view of their numerical results on violation of the ZG inequality (\ref%
{27}) by two-qudit states of a dimension $d$ in a range from $2$ to $10^{6}$
(see Fig. 1 in Ref. 31), Zohren and Gill conjectured\cite{31} that, for the
infinite dimensional \emph{optimal} bipartite states $\tau _{d,2},$ $%
d\rightarrow \infty ,$ specified by Fig 2 in Ref. 31, \emph{the\ tight
quantum analog} of the ZG inequality $\mathcal{B}_{zg}|_{_{_{_{lhv}}}}\geq 1$
under projective bipartite joint quantum measurements has the form\cite{32} 
\begin{equation}
\mathcal{B}_{zg}|_{\tau _{_{d,2}},\text{ }d\rightarrow \infty }\geq 0.
\label{29-1}
\end{equation}%
On the other hand, from (\ref{26}), (\ref{27}) it follows 
\begin{equation}
\mathcal{B}_{zg}|_{\tau _{d,2}}\geq \frac{3-\mathrm{\Upsilon }_{2\times
2}^{(\tau _{d,2})}}{2}.  \label{29}
\end{equation}%
This and the tightness for $d\rightarrow \infty $ of the quantum analog (\ref%
{29-1}), proved via the numerical results in Ref. 31, imply that, under
projective bipartite joint quantum measurements on an optimal state $\tau
_{d,2},d\rightarrow \infty $, 
\begin{equation}
0\geq 3-\mathrm{\Upsilon }_{2\times 2}^{(\tau _{d,2})}|_{d\rightarrow \infty
}\text{ \ \ }\Leftrightarrow \text{ \ \ }\mathrm{\Upsilon }_{2\times
2}^{(\tau _{d,2})}|_{d\rightarrow \infty }\geq 3.  \label{30}
\end{equation}%
However, for $N=S=2$, $d\rightarrow \infty ,$ the upper bound (\ref{24})
under projective measurements reads $\mathrm{\Upsilon }_{2\times 2}^{(\tau
_{d,2})}\mid _{d\rightarrow \infty }\leq 3.$ This and relation (\ref{30})
imply 
\begin{equation}
\mathrm{\Upsilon }_{2\times 2}^{(\tau _{d,2})}|_{d\rightarrow \infty }=3.
\label{31}
\end{equation}

In view of definition (\ref{1.15}) of the maximal violation parameter $%
\mathrm{\Upsilon }_{2\times 2}^{(\tau _{d,2})}$, this means that there must
exist a general $2\times 2$-setting Bell inequality where, under projective
bipartite quantum measurements on an optimal state $\tau _{d,2},d\rightarrow
\infty $, the term $3$ in the upper bound $\min \{ \sqrt{d},3\}$ in (\ref{24}%
) is attained, exactly or almost. Finding such a general Bell inequality is
a problem for a future research.

In conclusion, for the maximal quantum violation of \emph{general} Bell
inequalities,\ we have derived a new precise upper bound (\ref{20}),
reducing for $N=2$ to our bipartite bound $(2d-1)$ in Eq. (64) of Ref. 9 and
essentially improving for $N\geq 3$ all the known precise upper bounds for
general multipartite Bell inequalities listed in item \textrm{(d)} of the
Introduction.

Via the upper bounds (\ref{23}), (\ref{24}), the new upper bound (\ref{20})
also essentially improves the known\cite{33} precise upper bounds on the
maximal quantum violation of general multipartite Bell inequalities for $S$
settings per site. For some $S,d$ and $N$ discussed above, the upper bounds (%
\ref{23}), (\ref{24}) are attainable.

\begin{acknowledgments}
We would like to thank Carlos Palazuelos for helpful discussions on the
results of the previous publications, and the anonymous referees for their
many useful comments which helped to improve the readability of this paper.
\end{acknowledgments}


\begin{thebibliography}{99}
\bibitem{1} On general Bell inequalities, see Section II.

\bibitem{2} B. S. Cirel'son, Quantum generalizations of Bell's inequality.
Letters in Math. Phys\emph{.} \textbf{4}, 93--100 (1980).

\bibitem{3} B. S. Tsirelson, Quantum analogues of the Bell inequalities. The
case of two spatially separated domains. J. Soviet Math. \textbf{36},
557--570 (1987).

\bibitem{4} R. F. Werner and M. M. Wolf, All multipartite Bell correlation
inequalities for two dichotomic observables per site. Phys. Rev. A \textbf{64%
}, 032112 (2001).

\bibitem{5} V. Scarani, N. Gisin, Spectral decomposition of Bell's operators
for qubits. J. of Physics A: Math. Gen. \textbf{34}, 6043--6053 (2001).

\bibitem{6} This follows from the definition of the Grothendieck's constant $%
K_{G}^{(\mathbb{R})}$\ and Theorem 2.1 in Ref. 3.

\bibitem{7} D. Perez-Garcia, M. M. Wolf, C. Palazuelos, I. Villanueva, M.
Junge, Unbounded violation of tripartite Bell inequalities. Commun. Math.
Phys. \textbf{279}, 455--486 (2008).

\bibitem{8} M. Junge and C. Palazuelos, Large violation of Bell inequalities
with low entanglement. Commun. Math. Phys\emph{.} \textbf{306,} 695--746
(2011).

\bibitem{9} E. R. Loubenets, Local quasi hidden variable modelling and
violations of Bell-type inequalities by a multipartite quantum state. J.
Math. Phys\emph{. }\textbf{53,} 022201 (2012).

\bibitem{10} J. Briet, T. Vidick. Explicit Lower and Upper Bounds on the
Entangled Value of Multiplayer XOR Games. Commun. Math. Phys. \textbf{321},
181--207 (2013).

\bibitem{11} C. Palazuelos, On the largest Bell violation attainable by a
quantum state. J. Funct. Analysis \textbf{267}, 1959--1985 (2014).

\bibitem{12} E. R. Loubenets, Context-invariant and Local Quasi Hidden
Variable (qHV) Modelling Versus Contextual and Nonlocal HV Modelling. Found.
Phys\emph{.} \textbf{45, }840--850 (2015).

\bibitem{13} E. R. Loubenets, On the existence of a local quasi hidden
variable (LqHV) model for each $N$-qudit state and the maximal quantum
violation of Bell inequalities. Intern. J. of Quantum Information \textbf{14}%
, 1640010 (2016).

\bibitem{14} C. Palazuelos and T. Vidick, Survey on Nonlocal Games and
Operator Space Theory. J. Math. Phys\emph{.} \textbf{57}, 015220 (2016).

\bibitem{15} M. Junge, T. Oikhberg, C. Palazuelos, Reducing the number of
questions in nonlocal games. J. Math. Phys. \textbf{57 }(10), 102203 (2016).

\bibitem{16} That is, Bell inequalities of an arbitrary type -- either for
correlation functions or for joint probabilities or of a more complicated
form, see in Section II.

\bibitem{17} Throughout the article, the term "precise bound" means that, in
this bound, no any universal constant is involved -- in contrast to some
bounds in Refs. 8, 10, 11, 14 defined up to universal constants.

\bibitem{18} E. R. Loubenets, Nonsignaling as the consistency condition for
local quasi-classical probability modeling of a general multipartite
correlation scenario. J. Phys. A: Math. Theor. \textbf{45,} 185306 (2012).

\bibitem{19} E. R. Loubenets, Context-invariant quasi hidden variable (qHV)
modelling of all joint von Neumann measurements for an arbitrary Hilbert
space. J. Math. Phys\emph{. }\textbf{56,} 032201 (2015).

\bibitem{20} For this notion, see Section II.

\bibitem{21} See Proposition 3 in Ref. 19.

\bibitem{22} E. R. Loubenets, Multipartite Bell-type inequalities for
arbitrary numbers of settings and outcomes per site. J. Phys. A: Math. Theor%
\emph{.} \textbf{41, }445304 (2008).

\bibitem{23} On the general framework for the probabilistic description of
multipartite correlation scenarios, see Ref. 24.

\bibitem{24} E. R. Loubenets, On the probabilistic description of a
multipartite correlation scenario with arbitrary numbers of settings and
outcomes per site. J. Phys. A: Math. Theor. \textbf{41,} 445303 (2008).

\bibitem{25} For the main statements on the LHV modelling of a general
multipartite correlation scenario, see section 4 in Ref. 24.

\bibitem{26} Here, the term \emph{a tight LHV constraint }means that, in the
LHV frame, the bounds established by this constraint cannot be improved. On
the difference between the terms \emph{a tight linear LHV constraint }and 
\emph{an extreme linear LHV constraint} in case of, for example, the LHV
constraints on a linear combination of correlation functions, see the end of
Section 2.1 in Ref. 22.

\bibitem{27} See Proposition 1 in Ref. 9 for an $N$-partite case and
Proposition 1 in Ref. 28 for a bipartite case.

\bibitem{28} E. R. Loubenets, Quantum states satisfying classical
probability constraints. Banach Center Publ. \textbf{73,} 325--337 (2006),
e-print arXiv:quant-ph/0406139.

\bibitem{29} See definition 2 in Section II of Ref. 9.

\bibitem{30} E. R. Loubenets, Full locality of a noisy state for $N\geq 3$
nonlocally entangled qudits. E-print arXiv:1611.06723 (2016).

\bibitem{31} S. Zohren and R. D. Gill, Maximal Violation of the
Collins-Gisin-Linden-Massar-Popescu Inequality for Infinite Dimensional
States. Phys. Rev. Lett. \textbf{100}, 120406 (2008).

\bibitem{32} See Eq. (8) in Ref. 31.

\bibitem{33} To our knowledge, for the maximal quantum violation $\mathrm{%
\Upsilon }_{S_{1}\times \cdots \times S_{N}}^{(\rho _{d,N})}$ by a state $%
\rho _{d,N}$ of general $S_{1}\times \cdots \times S_{N}$-setting Bell
inequalities, the precise upper bounds are presented by Eq. (62) in Ref. 9
and by Eq. (19) of Ref. 13. The bipartite upper bound $\mathrm{\Upsilon }%
_{S\times S}^{(\rho _{d,2})}\leq C\min \{d,S\}$ presented by Eq.(01) in Ref.
15 (also, in Refs. 8, 14) is defined up to a universal constant. Note also
that the upper bounds in Refs. 10, 14 on the maximal quantum violation of
general Bell inequalities used in nonlocal games do not need to hold for all
general Bell inequalities.
\end{thebibliography}
\end{document}